\documentclass[reprint, amssymb, amsmath,aip,cha,]{revtex4-1}

\usepackage{graphicx}
\usepackage{amssymb}
\usepackage{epstopdf}
\usepackage{upgreek}
\usepackage{dcolumn}
\usepackage{bm}
\usepackage[final]{pdfpages}

\expandafter\ifx\csname package@font\endcsname\relax\else
 \expandafter\expandafter
 \expandafter\usepackage
 \expandafter\expandafter
 \expandafter{\csname package@font\endcsname}%
\fi

\newcommand{\be}{\begin{eqnarray}}
\newcommand{\ee}{\end{eqnarray}}
\newcommand{\bfig}{\begin{figure}}
\newcommand{\efig}{\end{figure}}

\newcommand{\units}[1]{\,\mathrm{#1}}
\newcommand*{\Perm}[2]{{}^{#1}\!P_{#2}}%

\begin{document}

\title{General purpose multiplexing device for cryogenic microwave systems}%

\author{Benjamin J. Chapman}
\email{benjamin.chapman@colorado.edu}
\affiliation{JILA, National Institute of Standards and Technology and the University of Colorado, Boulder, Colorado 80309, USA}
\affiliation{Department of Physics, University of Colorado, Boulder, Colorado 80309, USA}
\author{Bradley A. Moores}
\affiliation{JILA, National Institute of Standards and Technology and the University of Colorado, Boulder, Colorado 80309, USA}
\affiliation{Department of Physics, University of Colorado, Boulder, Colorado 80309, USA}
\author{Eric I. Rosenthal}
\affiliation{JILA, National Institute of Standards and Technology and the University of Colorado, Boulder, Colorado 80309, USA}
\affiliation{Department of Physics, University of Colorado, Boulder, Colorado 80309, USA}
\author{Joseph Kerckhoff}
\altaffiliation{Current address: HRL Laboratories, LLC, Malibu, CA 90265, USA}
\affiliation{JILA, National Institute of Standards and Technology and the University of Colorado, Boulder, Colorado 80309, USA}
\affiliation{Department of Physics, University of Colorado, Boulder, Colorado 80309, USA}
\author{K. W. Lehnert}
\affiliation{JILA, National Institute of Standards and Technology and the University of Colorado, Boulder, Colorado 80309, USA}
\affiliation{Department of Physics, University of Colorado, Boulder, Colorado 80309, USA}

\begin{abstract}
We introduce and experimentally characterize a general purpose device for signal processing in circuit quantum electrodynamics systems.  The device is a broadband two-port microwave circuit element with three modes of operation: it can transmit, reflect, or invert incident signals between 4 and 8 GHz.  
This property makes it a versatile tool for lossless signal processing at cryogenic temperatures.  In particular, rapid switching (${\leq} 15\units{ns}$) between these operation modes enables several multiplexing readout protocols for superconducting qubits.
We report the device's performance in a two-channel code domain multiplexing demonstration.  The multiplexed data are recovered with fast readout times (up to $400 \units{ns}$) and infidelities ${\leq} 10^{-2}$  for probe powers ${\geq} 7\units{fW}$, in agreement with the expectation for binary signaling with Gaussian noise.  
\end{abstract}


\maketitle
Superconducting qubits have recently emerged as a leading candidate for quantum information processing~\cite{devoret:2013}. These devices are straight-forward to fabricate~\cite{paik:2011} and exhibit coherence times on the order of $100 \units{\mu s}$ when embedded in 3D microwave cavities~\cite{jin:2015}. 
Furthermore, with quantum-limited amplifiers~\cite{castellanos:2008,abdo:2011,macklin:2015}, quantum non-demolition readout can be performed with fidelities exceeding 97\%~\cite{hatridge:2013}. Despite these promising characteristics of individual qubits, one major obstacle to scaling up quantum information processing is the requisite classical hardware. 

Within the circuit quantum electrodynamics architecture, the state of a qubit (or several qubits) can be encoded in the phase of a microwave tone transmitted through a cavity that contains the qubit(s). Qubit states may then be measured by detecting the cavity's transmission. This readout method requires a microwave receiver consisting of multiple circulators, a directional coupler, a cryogenic amplifier, warm amplifiers, and mixers~\cite{rigetti:2012}.  Consequently, each cavity readout requires a copy of this bulky, power hungry, and expensive measurement chain.  

Multiplexing is a conventional solution to such a scaling challenge.  In addition to its widespread implementation in communication networks, multiplexing techniques are applied broadly in low-temperature physics.  For example, in the last two decades multiplexing has rapidly accelerated the readout of detector arrays used in astrophysics, high energy, and materials science~\cite{benford:2000,yoon:2001,doriese:2004,lanting:2005,benoit:2008,mates:2008,irwin:2010,irwin:2012}. 

The readout of many superconducting-qubits with frequency domain multiplexing has recently been demonstrated as a viable scheme for reducing hardware overhead~\cite{chen:2012,jerger:2012,barends:2014,riste:2015}.  However, some applications (e.g. quantum simulators) require nearly identical qubit/cavity systems.  
In these cases, other multiplexing techniques such as time and code domain schemes~\cite{couch:1993} are favorable. 

In code domain multiplexing, each channel is modulated by distinct elements from an orthogonal set of functions.  The orthogonality of the set is then exploited to recover the channel's original signal.  This combines attractive features of frequency and time domain multiplexing, allowing for simultaneous measurement of all channels while maintaining the ability to dynamically allocate bandwidth.  These considerations reflect the fact that in any practical application, hardware constraints typically require a combination of frequency, time, and code domain techniques to maximize bandwidth. 

To enable a hybrid multiplexing approach for qubit readout, here we introduce a multipurpose
device that we call a Tunable Inductor Bridge (TIB). Each TIB can be tuned to either transmit, reflect, or invert a microwave tone (see Fig.~\ref{fig:fig1}(a)). Rapidly alternating between the transmit/reflect or transmit/invert operation modes allows the TIB to act as a fast switch or phase-chopper.  This dual functionality makes the device ideal for implementing both time and code domain schemes, as shown in Fig.~\ref{fig:fig1}(b). For example, sequential qubit manipulation and readout can be implemented by positioning fast-switching TIBs upstream from each qubit. Alternatively, qubits can be simultaneously readout in a code domain scheme by positioning TIBs downstream from each qubit/cavity system, and operating them as phase-choppers to spread the cavity output spectrum (beyond the qubit/cavity bandwidth).  
In this letter, we describe the design, layout, and theory of operation for the TIB.  We characterize it experimentally, and present a two-channel code domain multiplexing demonstration.

\begin{figure}[htb]
\begin{center}
\includegraphics[width=1.0\linewidth]{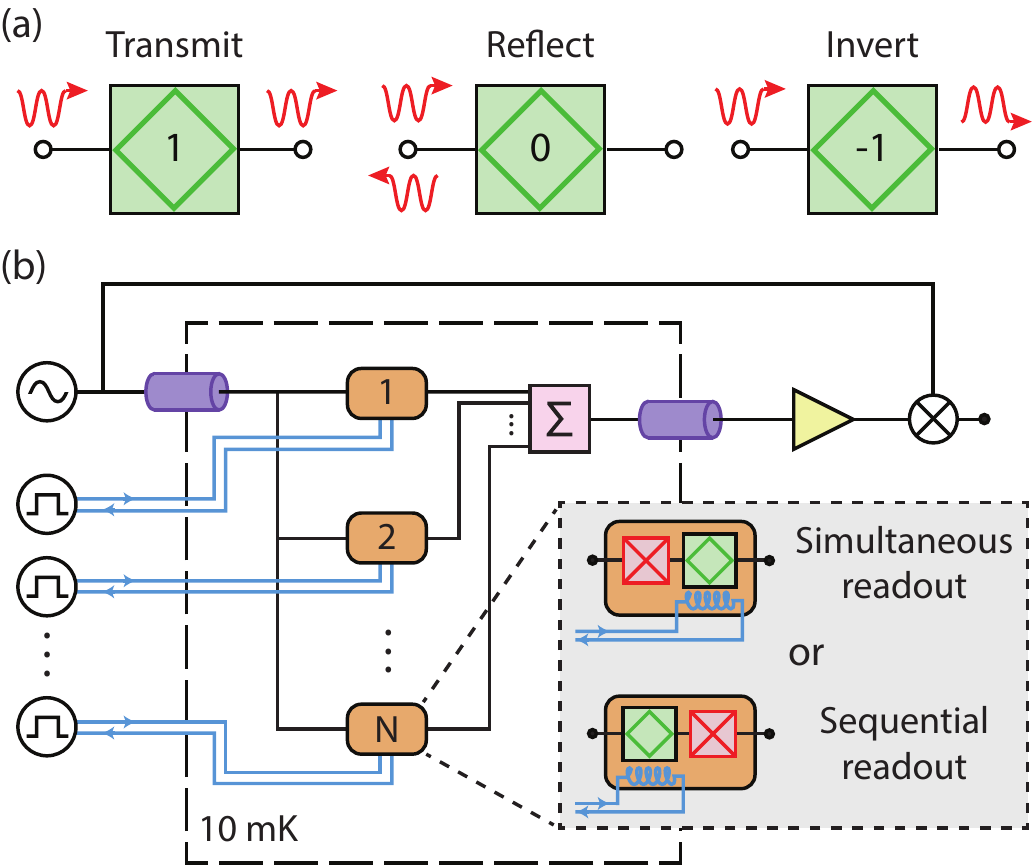}
\caption {\small (a) The three operation modes of a Tunable Inductor Bridge (TIB): transmit, reflect, and invert.  (b) Conceptual schematic for the proposed multiplexed readout of $N$ superconducting qubit/cavity systems (red x's) using $N$ TIBs (green diamonds).  Modification of the TIBs' position in the wiring allows implementation of either time 
or code 
domain multiplexing.  Readout requires only a single microwave receiver; additional isolators are not needed in the $N$ multiplexed channels.  Recombination of the multiplexed signals (indicated by the summation box $\Sigma$) may be accomplished with a microwave hybrid, as in the demonstration in this letter, or by working in a lumped-element limit, which avoids the loss of a matched recombination network~\cite{supp}.
}
\label{fig:fig1}
\end{center}
\efig

\begin{figure}[htb]
\begin{center}
\includegraphics[width=1.0\linewidth]{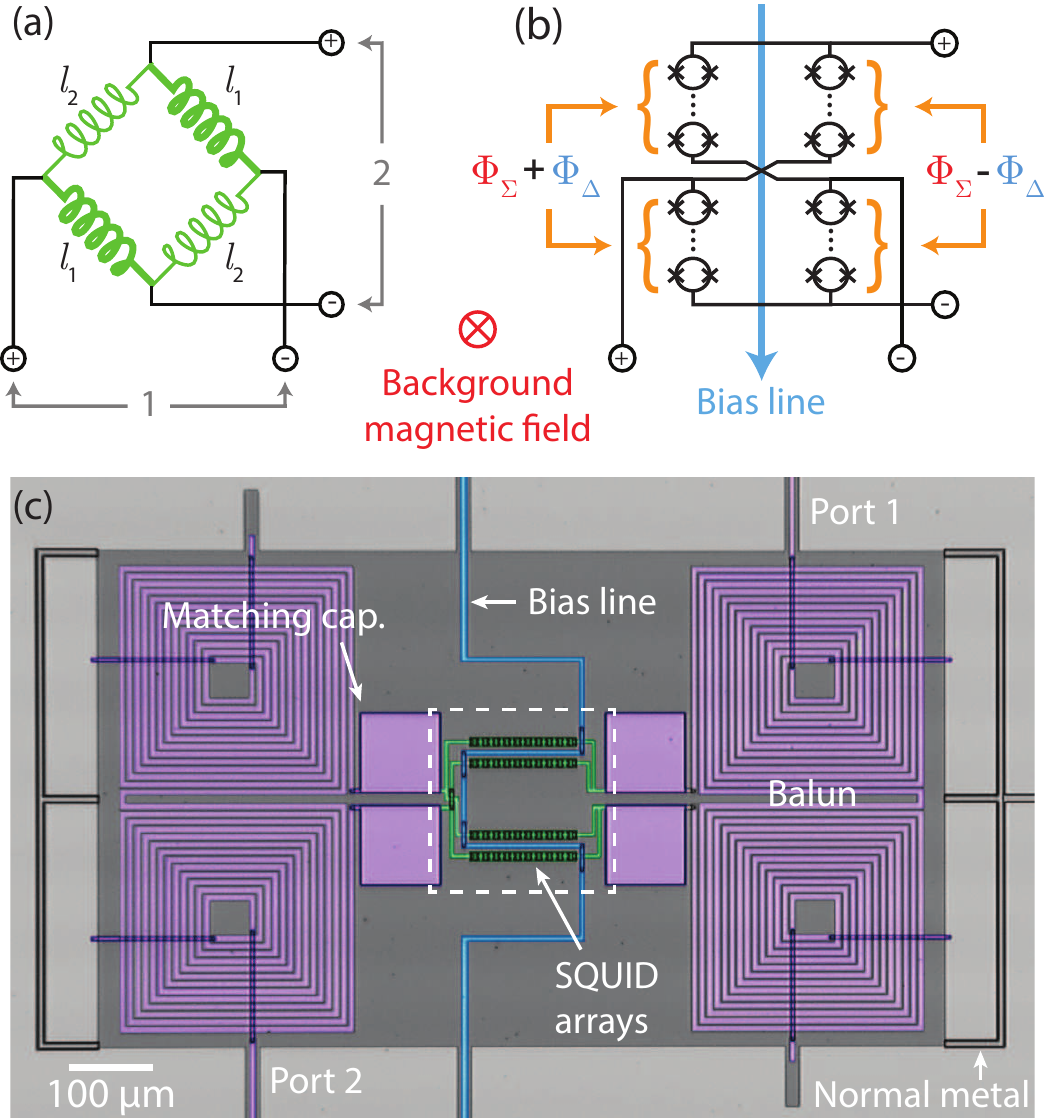}
\caption {\small (a) Lumped element schematic of the TIB, formed with 
a bridge of tunable inductors.  (b) Layout of a TIB 
using series arrays of SQUIDs as inductors.  Modulation of the inductors is accomplished with an off-chip magnetic coil 
and an on-chip bias line.  
(c) False-color photo of the fabricated chip.  Capacitors (purple) match the network to 50 $\Omega$ and break superconducting loops in the circuit.  Marchand baluns~\cite{marchand:1944} (purple) allow the four bridge nodes to be driven as differential ports.  Additional superconducting loops are broken by normal metal in the ground plane (light gray).  The bridge of SQUID-array inductors (green) is tuned with the on-chip bias line (blue).  A dashed white box indicates how the conceptual layout in (b) is embedded in the circuit.  Devices were fabricated at NIST Boulder in a NbAlO$_x$Nb tri-layer process\cite{mates:2008}. 
}
\label{fig:fig2}
\end{center}
\efig

The TIB is a two-port microwave device composed of four tunable inductors arranged as a Wheatstone bridge 
(see Fig.~\ref{fig:fig2}(a)).  The four inductors are split into two pairs that tune in tandem.  
Current in an on-chip bias line changes the inductance of the pairs in opposite directions, imbalancing the bridge.  Examination of the transmission coefficient $T$ reveals how changing this imbalance adjusts the TIB between its 
operation modes.  When coupled to transmission lines of characteristic impedance $Z_0$, the forward scattering parameter at angular frequency $\omega$ is
\be
T = \frac{i \omega  (l_1 - l_2) Z_0}{(i \omega l_1  + Z_0) (i  \omega l_2 + Z_0)}.
\label{T}
\ee
Transmission is clearly nulled when the inductors $l_1$ and $l_2$ are equal, realizing the device's reflect operation mode. Switching to the transmit mode is accomplished by maximally imbalancing the bridge.  Lastly, reversing the sense of this imbalance inverts the transmitted signal, as $T$ is odd under exchange of $l_1$ and $l_2$.

To realize tunable inductors for cryogenic microwave applications, we use series arrays of superconducting quantum interference devices (SQUIDs).  When the geometric inductance of the SQUIDs is small with respect to their Josephson inductance, the critical current $I_s$ of these SQUIDs tunes with the magnetic flux $\Phi$ that threads through them as
\be
I_s = 2 I_0 \left\lvert \cos{\left( \frac{\Phi}{2 \phi_0}\right)}\right\rvert .
\label{SQUIDIC}
\ee
Here $\phi_0 = \hbar/2e$ is the reduced flux quantum and $I_0$ is the critical current of the Josephson junctions.
When the current flowing through the arrays is small compared to $I_s$, the array inductance is
\be
l = N_\textrm{sq} \frac{\phi_0}{I_s},
\label{arrayl}
\ee
where $N_\textrm{sq}$ is the number of SQUIDs in the series array, and all junctions are assumed to be identical.  

To layout a bridge of tunable inductors using SQUID arrays, we use a previously proposed figure-eight geometry~\cite{kerckhoff:2015}. The simultaneous tuning of the inductor pairs is accomplished in two steps (depicted in Fig.~\ref{fig:fig2}(b)).  First, an off-chip coil creates a background magnetic field of uniform strength across the chip.  This threads a magnetic flux $\Phi_\Sigma$ through all the SQUIDs, while the gradiometric layout of the figure-eight ensures that no net flux pierces the bridge.   Second, an on-chip bias line carries a current that simultaneously threads a flux $\Phi_\Delta$ through the two arrays on one side of the line, and a flux $-\Phi_\Delta$ through the other two arrays.  Each SQUID in one pair of arrays is pierced by a sum of magnetic fluxes from the background coil and the bias line $\Phi_\Sigma + \Phi_{\Delta}$, while SQUIDs in the other pair of arrays are pierced by a total flux $\Phi_\Sigma - \Phi_{\Delta}$.  These differing fluxes result in different critical currents, as given in Eq.~\ref{SQUIDIC}, and hence different inductances, as given in Eq.~\ref{arrayl}. The figure-eight ensures the connectivity of the inductors corresponds to Fig.~\ref{fig:fig2}(a), with each array opposite its equal in the bridge.  A false-color image of the fabricated device is shown in Fig.~\ref{fig:fig2}(c).

To assess the TIB's performance as a phase-chopper, we performed a homodyne measurement on a microwave signal transmitted through the bridge.  A schematic of the measurement is shown in Fig.~\ref{fig:fig3}(a).  During the measurement, a small current in the TIB's bias line ($< 200 \units{\mu A}$) is modulated to tune the device between its transmit and invert modes.  The resulting mixed-down voltage is then digitized and shown in Fig.~\ref{fig:fig3}(b).  
Each trace shows transmission with a different bias line modulation, in which the operation mode was switched 1, 2, 3, 4, or 64 times during the measurement period of $10\units{\mu s}$.


\begin{figure}[htb]
\begin{center}
\includegraphics[width=1.0\linewidth]{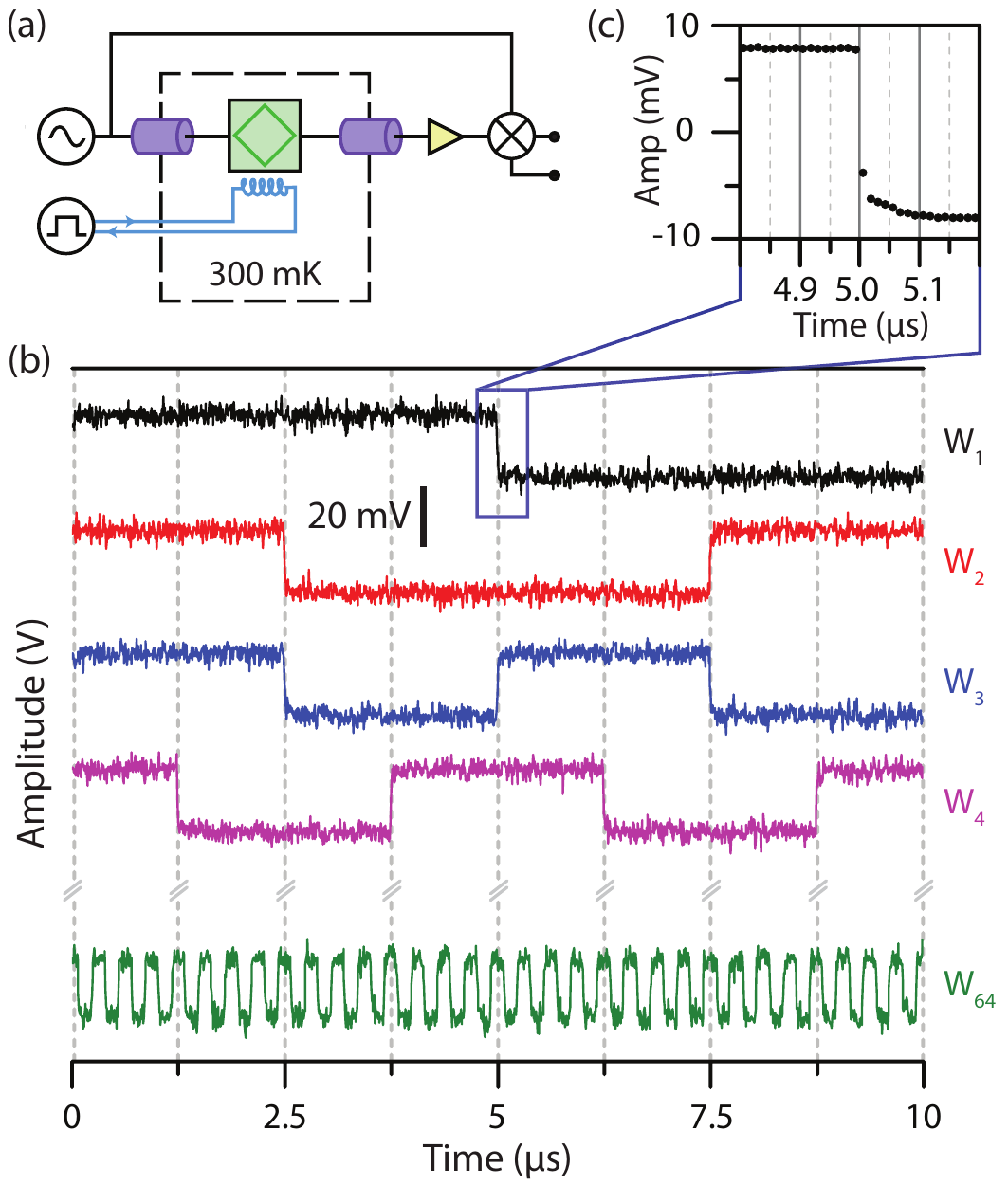}
\caption {\small (a) Schematic of the homodyne measurement used to demonstrate the TIB's performance as a phase-chopper.  A low-bandwidth control line (blue coil) is used to switch the TIB between its transmit and invert modes.  (b) Measured single-shot homodyne voltage traces, phase-chopped by the TIB.  The current in the bias line of the TIB was modulated to create five elements in a set of orthogonal functions known as Walsh codes.  (c) Magnification of the falling edge in $W_1(t)$, averaged 1280 times.  The switching time is $15 \units{ns}$ (sampling interval is $12.5 \units{ns}$).}  
\label{fig:fig3}
\end{center}
\efig

Switching between the transmit and invert modes occurs rapidly. Fig.~\ref{fig:fig3}(c) shows one such falling-edge in finer time-resolution. The observed switching time of $15 \units{ns}$ (sampling interval time is $12.5 \units{ns}$) is limited by an $80 \units{MHz}$ low-pass filter after the mixer in the receiver chain~\cite{supp}.  In principle, switching times on the order of 1 ns could be expected, constrained by the bandwidth of the Marchand baluns and the bandwidth over which the capacitors match the network to $Z_0$.

In addition to the switching time, other relevant specifications of the TIB are its linearity, on-off ratio, insertion loss, and phase balance~\cite{supp}.  The $N_\textrm{sq} = 20$ SQUIDs in each array form inductors with high-power handling~\cite{kerckhoff:2015}.  This is reflected in power-sweeps of the scattering parameters, which are linear up to powers of about $1 \units{pW}$ in the three tested devices.  For reference, dispersive-readout typically uses microwave tones with powers less than $1 \units{fW}$~\cite{riste:2013}.  The on-off ratio (the ratio of the transmission coefficients in the transmit and reflect operation modes) can be tuned 
above $20 \units{dB}$ over the entire $4\mbox{--}8\units{GHz}$ range, and  
$40\units{dB}$ at the designed center frequency of $6\units{GHz}$.  The devices are also low-loss: the two TIBs used for the multiplexing demonstration in this letter have insertion losses below $0.5 \units{dB}$.  These specifications compare favorably with other recent realizations of fast, Josephson-junction based switches~\cite{naaman:2015}.  Finally, the average magnitude of the phase imbalance between the transmit and invert modes is 5 degrees over the $4\mbox{--}8 \units{GHz}$ band.

A multiplexed readout of an $N$-qubit/cavity system is beyond the scope of this paper.  To illustrate a proof of concept, here we present a multiplexed readout of an analogous two-channel system arranged to simulate the microwave signals that would be generated by periodically measuring two qubit/cavity systems simultaneously.  

To modulate the channels in this code domain demonstration, two TIBs are programmed to switch between their transmit and invert modes according to distinct elements in the orthogonal and periodic (period $t_w$) set of functions $\{w_n\}$ known as Walsh codes~\cite{walsh:1923}.  As the TIBs have finite bandwidth, modulation of the channels is not instantaneous.  We denote our finite-bandwidth experimental realization of the Walsh codes as $\{W_n\}$, to distinguish them from the mathematical set $\{w_n\}$~\cite{Wm}.  
The traces shown in Fig.~\ref{fig:fig3}(b) are five examples of elements in $\{W_n\}$ with $t_w = 10\units{\mu s}$.  

In Fig.~\ref{fig:fig4}(a), a schematic of our homodyne measurement shows how two Walsh-modulated channels may be code domain multiplexed.  
First, we generate a string of $p$ pseudorandom digital bits $\bm{D_1}$ with an arbitrary waveform generator, and use these to modulate the phase of a 5.88 GHz microwave tone.  The resulting waveform consists of a tone whose phase jumps pseudorandomly in time between $0$ and $\pi$.  This arrangement is duplicated with a second sequence of random digital voltages $\bm{D_2}$ to simulate the output of a second qubit/cavity system~\cite{theta}.

\begin{figure}[htb]
\begin{center}
\includegraphics[width=1.0\linewidth]{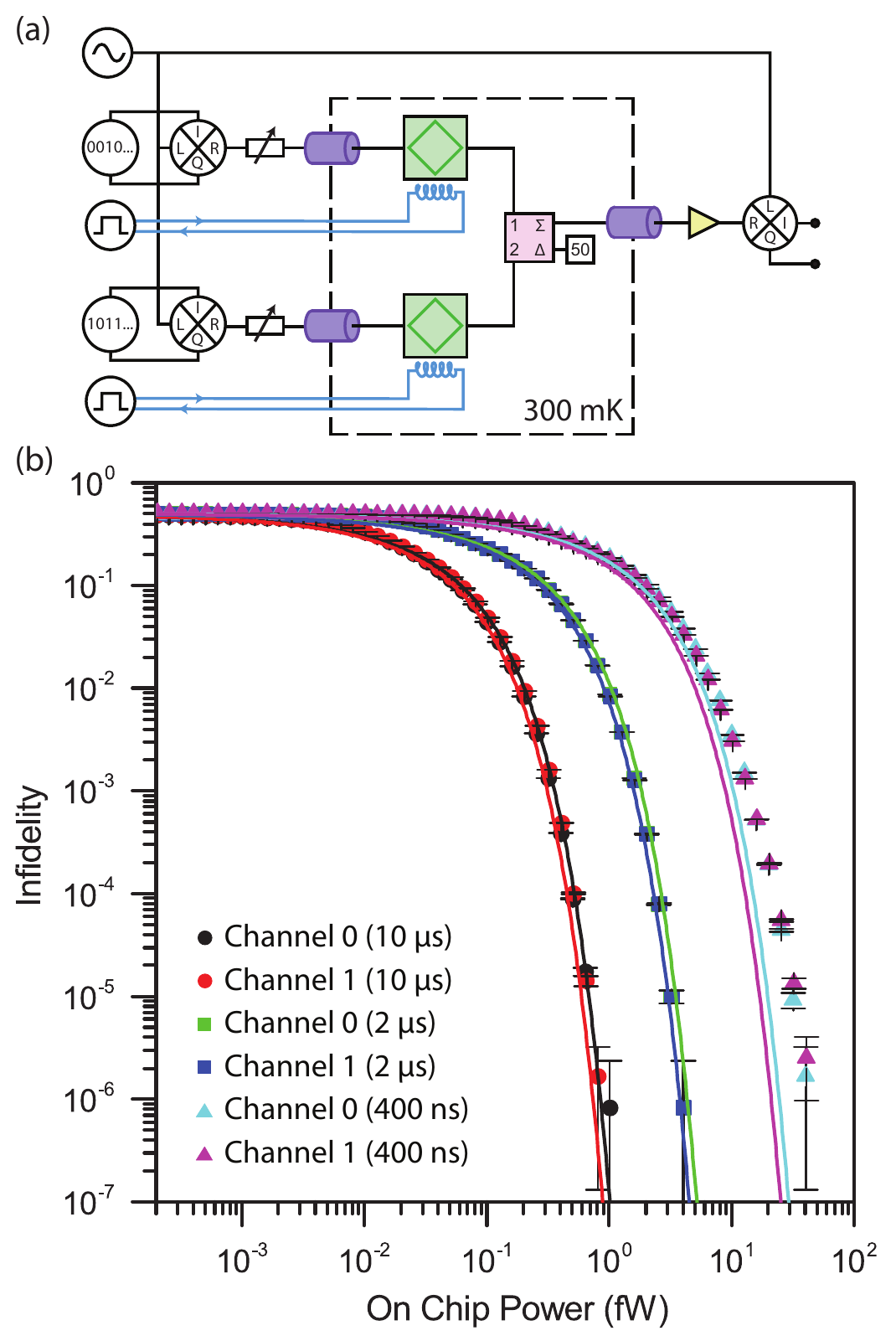}
\caption {\small (a) Schematic of the homodyne measurement used to demonstrate two-channel code domain multiplexing.
Two random streams of digital bits are encoded in the phase of microwave carriers and passed into a $^3$He cryostat, where they are modulated by two TIBs and then summed and readout in a single measurement chain.  Both TIBs are on the same chip.  The measured voltage is compared with the modulation applied by each TIB to reconstruct the original data streams.  (b) Readout infidelity as a function of microwave signal power for both channels and three different bit transmission-durations: $t_w = 400 \units{ns}$, $2 \units{\mu s}$, and $10 \units{\mu s}$.  Solid lines are the prediction of a Gaussian noise model with no adjustable parameters~\cite{supp}.}
\label{fig:fig4}
\end{center}
\efig

These microwave channels are then fed into a cryostat and modulated by separate TIBs. During one digital period, the two TIBs are programmed to 
transmit and invert according to distinct Walsh codes $w_n$ and $w_m$.  The outputs of the TIBs are summed and directed out of the cryostat through a single microwave receiver.  The final mixed down voltage is the sum of two Walsh-modulated analog square waves, where the upper and lower values of the two square waves correspond to the logical 1 and 0 of the original digital bits. We call the (mixed-down) analog versions of the two bit streams $\bm{A_1}$ and $\bm{A_2}$.  

The time trace of the $k^{\textrm{th}}$ bit takes the form
\be
\label{recon}
V_k(t) = A_{1,k} |T_{\textrm{ch}1}| W_n(t) + A_{2,k} |T_{\textrm{ch}2}| W_m(t).
\ee
Here $T_{\textrm{ch}1}$ ($T_{\textrm{ch}2}$) is the transmission coefficient of the TIB in channel 1 (2).  Reconstruction $A^r_{q,k}$ of the $k^{\textrm{th}}$ bit in the $q^{\textrm{th}}$ channel follows from the orthogonality of the Walsh codes~\cite{nonorthogonality}.    
For example,
\be
A^r_{1,k} = \frac{1}{t_w}\int_{(k-1)t_w}^{kt_w} V_k(t') W_n(t') dt'.
\ee
This analog reconstruction may then be digitized with a thresholding procedure and compared to the transmitted digital bit to determine if a fault occurred.  

To characterize our two-channel multiplexing demonstration, we ran the above protocol with $p = 10^5$ digital bits passing through each channel.  
We then repeated that process for the $\Perm{4}{2} = 12$ permutations for assigning Walsh codes from the set $\left\{W_1,W_2,W_3,W_4\right\}$ to a pair of channels.  For each realization of the protocol, we recorded the number of faults in the reconstruction of each channel, and divided this by the total number of transmitted bits to obtain a channel infidelity.  Fig.~\ref{fig:fig4}(b) shows the infidelity in both channels as a function of the microwave power delivered to the TIBs~\cite{avg}.  Three different bit durations of $t_w = 400 \units{ns}$, $2 \units{\mu s}$ and $10 \units{\mu s}$ are shown, with error bars indicating 95\% confidence intervals, calculated for a binomial distribution. Solid lines are 
the 
predictions for a binary signaling process~\cite{madhow:2014}.  The predictions have no adjustable parameters and are made from the measured spectral densities of the signal and noise in each channel~\cite{supp}.

We observe that the readout infidelities are independent of the permutation of Walsh codes for $t_w>1 \units{\mu s}$.  For $t_w \leq 1 \units{\mu s}$, timing considerations such as the relative delay between the drive and control lines and the imperfect orthogonality of ${W_n}$ become relevant.  
These factors cause cross-talk and intersymbol interference which lead to performance differences between the permutations.  They also cause a discrepancy between the model's predictions and the measured infidelities, especially when the predicted infidelity is less than $10^{-3}$.  Fidelity may be improved in this regime with standard methods like pulse-shaping, equalization, and frame synchronization~\cite{couch:1993}.

In summary, we present here an experimental demonstration of a broadband tri-state device that can rapidly (${\leq}15\units{ns}$) switch between its transmit, reflect, and invert operation modes.  The device is realized with a bridge of inductors built from SQUID arrays, making it a multipurpose tool for implementing time and code domain multiplexing schemes in a cryogenic microwave environment.
To illustrate its capabilities, we have demonstrated a two-channel code domain multiplexed readout of a pair of random bit streams.  With a $400 \units{ns}$ readout time (compatible with demonstrated qubit readout times~\cite{riste:2013,hatridge:2013,murch:2013,riste:2015,chen:2016}),  
infidelities ${\leq} 10^{-2}$ are achieved when the power entering the TIBs exceeds $7\units{fW}$.  Although this power is larger than the $\sim0.3$ fW used in typical qubit readout~\cite{riste:2013}, 
our HEMT amplifier is about 20 times noisier than a quantum limited amplifier. To the degree that the measured infidelity is accurately predicted by a binary signaling model, the code domain scheme will not decrease qubit readout fidelity.  
This demonstration represents a step toward a flexible and scalable read-out architecture for superconducting qubits.

\vspace{0.1in}
\noindent{\emph {Acknowledgment}} This work is supported by the ARO under contract W911NF-14-1-0079 and the National Science Foundation under Grant Number
1125844.


\begin{thebibliography}{36}
\bibitem [{Wm()}]{Wm}%
  \BibitemOpen
  \href@noop {} {\ }\bibinfo {note} {Unlike $\{w_n\}$, the set $\{W_n\}$ is
  only approximately orthogonal, with $|\frac{1}{T_w} \int_0^{T_w} W_n(t)
  W_m(t)dt| = \epsilon \ll 1$ when $n \neq m$. We find $\epsilon \leq 3 \times
  10^{-2}$ when $T_w \geq 400 \units{ns}$.}\BibitemShut {Stop}%
\bibitem [{avg()}]{avg}%
  \BibitemOpen
  \href@noop {} {\ }\bibinfo {note} {The displayed infidelities are averaged
  over the 12 Walsh code permutations.}\BibitemShut {Stop}%
\end{thebibliography}
%

\pagebreak
\widetext
\begin{center}
\textbf{\large Supplementary Material for\\ ``General purpose multiplexing device for cryogenic microwave systems''}
\end{center}
\setcounter{equation}{0}
\setcounter{figure}{0}
\setcounter{table}{0}
\setcounter{page}{1}
\makeatletter
\renewcommand{\theequation}{S\arabic{equation}}
\renewcommand{\thefigure}{S\arabic{figure}}
\renewcommand{\bibnumfmt}[1]{[S#1]}
\renewcommand{\citenumfont}[1]{S#1}

\section{Experimental setup for multiplexing demonstration}

The experimental setup for the code domain multiplexing demonstration shown in Fig. 4 is depicted in Fig.~\ref{fig:supp1}. Two pseudorandom bit streams, $D_1$ and $D_2$, are created and loaded into an arbitrary waveform generator. The outputs of the generator are connected to the I and Q ports of two mixers driven by the same local oscillator.  This results in two microwave signals whose phase is periodically but psuedorandomly shifted by $\pi$, in a pattern determined by the sequences $D_1$ and $D_2$.  These tones then enter a $^3$He cryostat, where each is passed into a separate TIB. 
An off-chip coil, represented by an orange circle, produces a static background magnetic field around both TIBs. On-chip bias lines, represented by the blue coils, tune the imbalance of the bridges. Together, these controls allow the TIBs to be used as phase-choppers, modulating the incident microwave tones with a Walsh code.  The output signals from the two TIBs are then summed using a 180-degree hybrid coupler, and readout in a microwave receiver.  Note that the demonstration does not model the detrimental effects of qubit decay on readout fidelity.

\begin{figure}[htb]
\begin{center}
\includegraphics[width=0.8\linewidth]{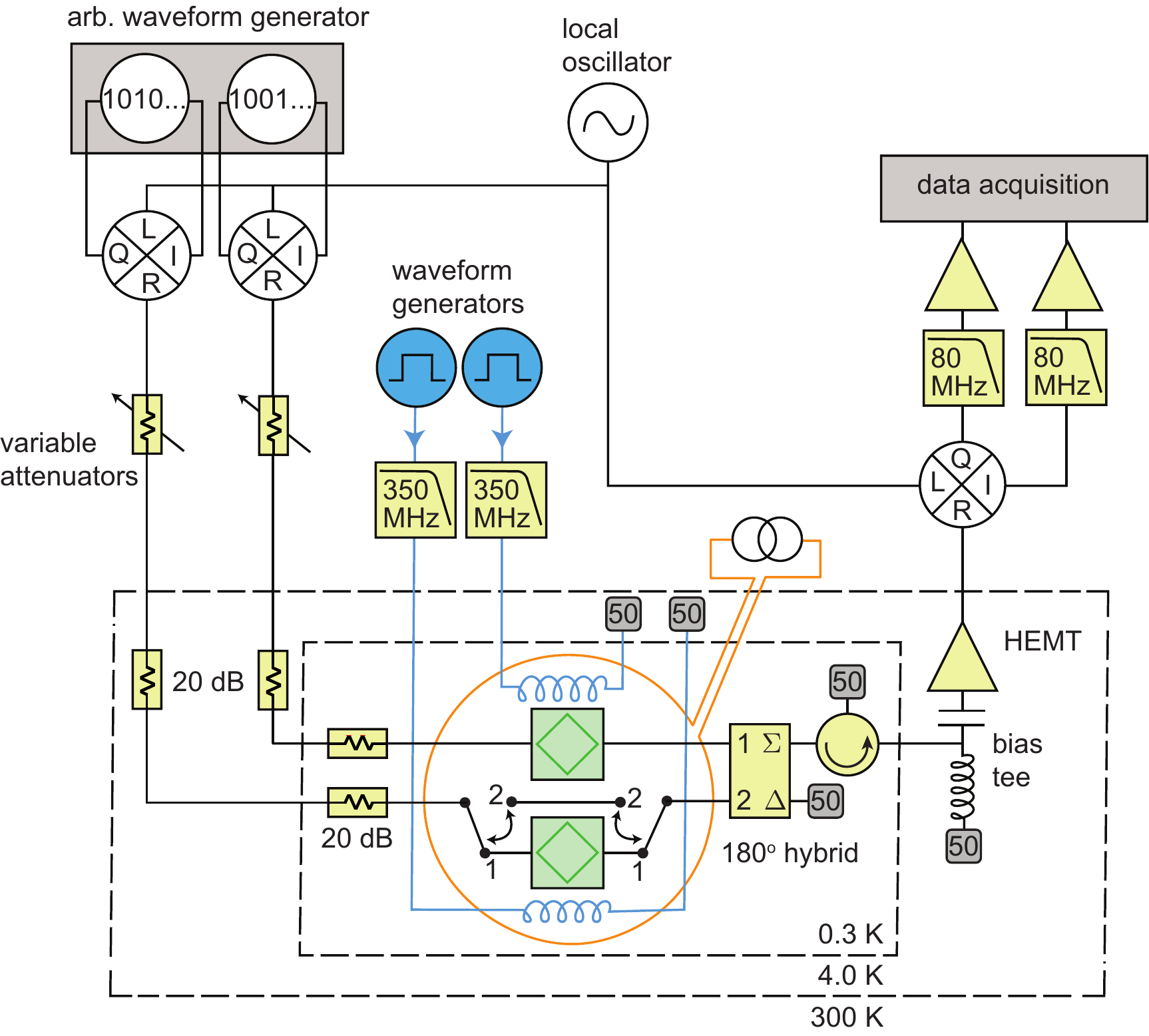}
\caption {\small Wiring diagram of the multiplexing demonstration.  The orange circle represents an off-chip magnetic coil.}
\label{fig:supp1}
\end{center}
\end{figure}

\section{Transmission of a TIB built from SQUID arrays}

Based on Eqs.~2 and~3 of the main text, the SQUID arrays of a TIB have inductance
\begin{equation} \label{eqn:supp1}
L = \frac{l}{|\cos \big(\frac{\Phi_{\Sigma} \pm \Phi_{\Delta}}{2 \phi_0} \big)|}.
\end{equation}
Here $\phi_0 = \hbar /2 e$ is a reduced flux quanta, and $l = \left(N_{sq}/2\right) l_J$ with $l_J = \phi_0/I_0$ the Josephson inductance  of a single junction.  For our devices, $N_{sq} = 20$ and the junction critical current $I_0 = 6.5$ $\units{\mu}$A, giving a maximum SQUID critical current of 13 $\mu$A and a minimum inductance of $l = 0.5$ nH.  We operate the devices by fixing the magnetic flux $\Phi_{\Sigma}$ from the off-chip coil. Tuning between the transmit, reflect, and invert operation modes is accomplished by varying $\Phi_{\Delta}$, the magnetic flux from the on-chip bias line.

The geometric inductance of the SQUID arrays (neglected in the expressions in the main text) limits the dynamic range over which the inductance of the arrays may be tuned.  This in turn limits the degree to which the bridge may be imbalanced.  Capacitors placed in series with the four nodes of the bridge (as shown in Fig.~2(c)) match the finitely-imbalanced bridge to $50 \units{\Omega}$.  This is manifest in reflection measurements, which show vanishing reflection in the transmit and invert operation modes.

Fig.~\ref{fig:supp2} shows the transmission of a TIB, as a function of frequency and on-chip bias flux $\Phi_\Delta$.  Measurements at three different values of the off-chip flux $\Phi_\Sigma$ are shown in (a-c).  For comparison, (d-f) show numerical simulations of the device computed with a planar method of moments solver. For the multiplexing demonstration in the main text, we operate with $\Phi_\Sigma = 0.13$ and a frequency of $5.88 \units{GHz}$.

\begin{figure}[htb]
\begin{center}
\includegraphics[width=0.9\linewidth]{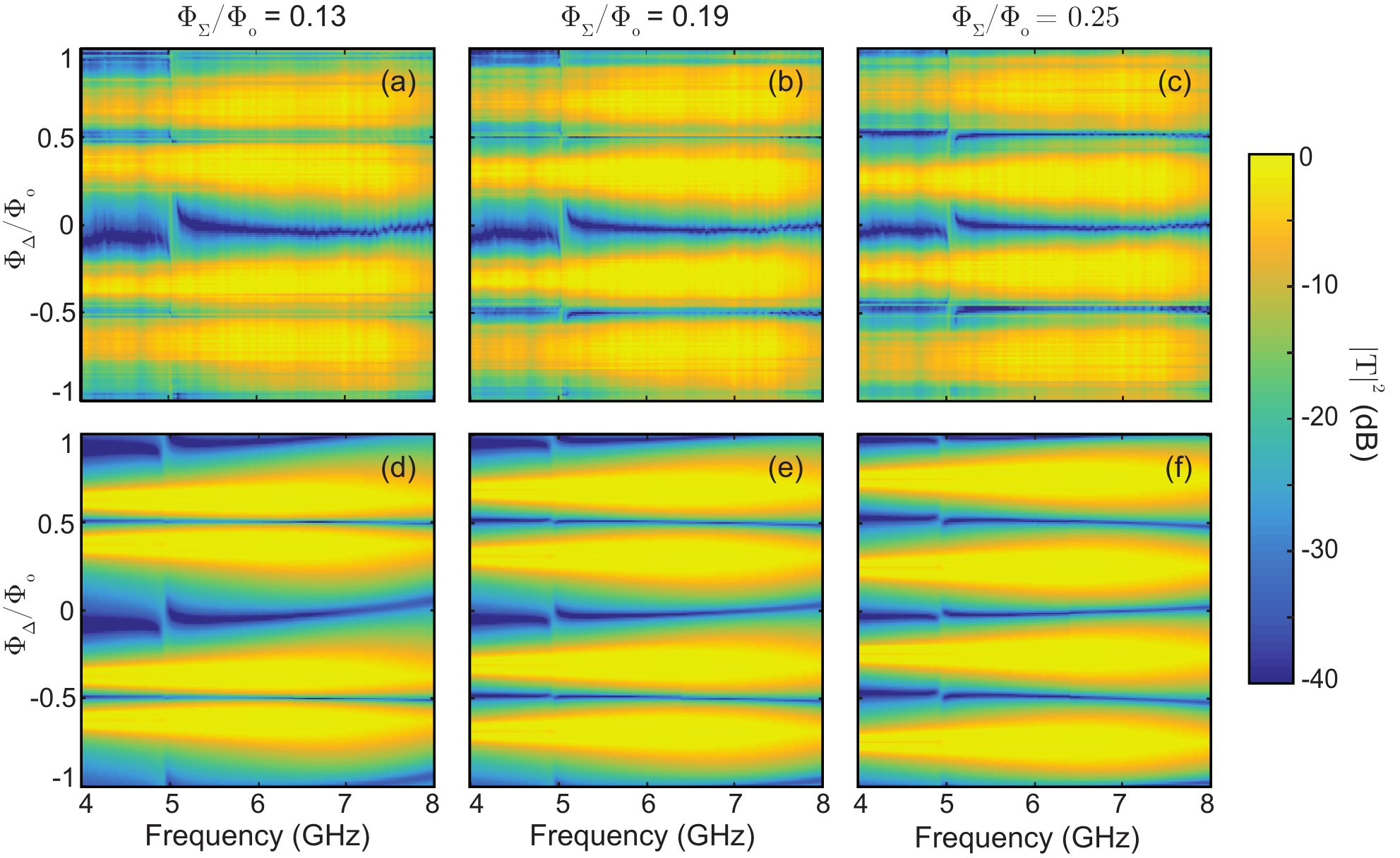}
\caption {\small The transmitted power $|T|^2$ of a TIB, as a function of frequency and on-chip bias flux $\Phi_\Delta$.  (a-c) Measurements at three different values of off-chip magnetic flux, $\Phi_\Sigma/\Phi_0 = 0.13$, 0.19, and 0.25, respectively. (d-f) Calculations of $|T|^2$ computed with a planar method of moments solver.}
\label{fig:supp2}
\end{center}
\end{figure}

The transmitted power in the transmit and invert modes is near unity, while the reflect mode suppresses transmission by over $20 \units{dB}$.  Furthermore, the operation bandwidth is large, spanning from 4--8 GHz.  Within this interval, the TIB can be operated at any frequency except in a narrow band around $5 \units{GHz}$, where a chip mode degrades performance. Since numerical simulations predict the chip mode, this notch can be removed in future designs by adjusting dimensions of the baluns.

\section{Device specifications}
Several specifications quantify the performance of a phase-chopper or fast-switch.  In addition to the switching time discussed in the main text, the on-off ratio of the switch, the phase balance of the phase-chopper, and their linearity and insertion loss determine their utility.  Fig.~\ref{fig:supp3}(a) shows the magnitude of the squared transmission coefficient at $5.88 \units{GHz}$ as a function of the applied power.  Transmission is independent of power below a critical value of $\approx 1$ pW.  The bypass switch illustrated in Fig.~\ref{fig:supp1} is used to calibrate the absolute magnitude of $|T|^2$ for TIB \#1.  TIB \#2 is calibrated with the same data, less the room temperature difference in the attenuation of their input lines scaled to 300 mK. The on-off ratio of a TIB operated as a fast-switch is shown in Fig.~\ref{fig:supp3}(b) (plotted traces are the best achievable performance at each frequency).  Performance is broadband, excluding a narrow notch around $5 \units{GHz}$.  Typical values exceed 40 dB.  Finally, the phase difference (less 180 degrees) is shown in Fig.~\ref{fig:supp3}(c).  Phase imbalance is typically less than 10 degrees over the operation band.

\begin{figure}[htb]
\begin{center}
\includegraphics[width=1\linewidth]{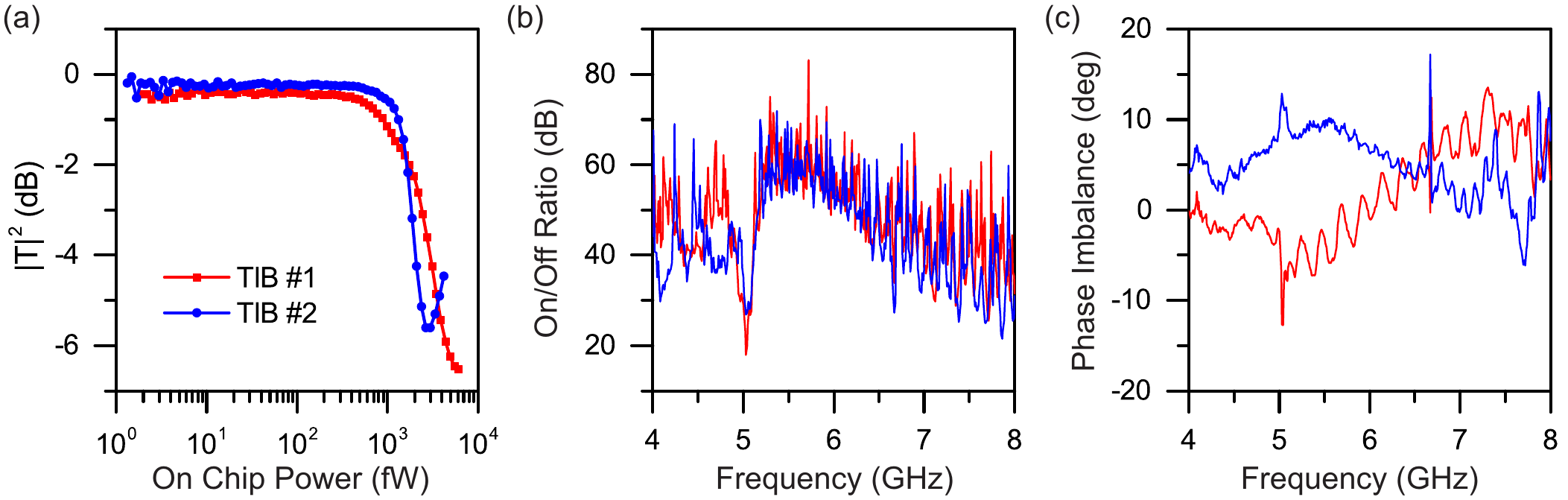}
\caption {\small (a) Transmitted power as a function of input power at 5.88 GHz.  The devices are linear up to approximately 1 pW.  (b) The (best achievable) ratio of transmitted power to reflected power when the TIBs are operated as fast switches, as a function of frequency.  For fixed flux control parameters, on-off ratios greater than 20 dB are achieved over the entire 4--8 GHz band, excepting a 30 MHz notch around the chip mode.  On-off ratios greater than 30 (40) dB have instantaneous bandwidths greater than 2 GHz (500 MHz). (c) Phase difference between the transmit and invert operation modes (less 180 degrees) when the TIBs are operated as phase-choppers.  Phase imbalance is less than 10 degrees over a broad range of frequencies.  In (a-c), red traces represent the performance of TIB \#1, and blue traces represent TIB \#2.  Both devices are on the same chip.}
\label{fig:supp3}
\end{center}
\end{figure}

\section{Power Recombination of Multiplexed Channels}
As discussed in the caption of Fig.~1 of the main text, microwave signals in the distinct qubit/cavity channels must be recombined for multiplexed readout. However, multiple incoherent signals propagating in separate transmission lines and at the same frequency cannot be combined into one transmission line without loss~\cite{pozar:2011}. At a matched $N$-port network each input line would deliver on-average $1/N^\textrm{th}$ of its power to the single output line. On the other hand, if each channel behaved as a stiff current source, all channels could be summed losslessly and without interference. In this scheme, all channels drive current into a single node with $N+1$ branches, where one branch is the 50 $\Omega$ receiver line. In effect, a phase-chopped cavity output acts as a stiff current source if the modulation frequency is large compared to the cavity linewidth. Using this strategy for code domain multiplexing requires either that the channels are spatially close to one another so that the connections between the cavity outputs and the common node are much shorter than a wavelength, or that these connections are chosen to be a half-wavelength long. 
For a time domain multiplexing architecture, lossless power combination may also be achieved at a node with $N +1$ branches by incorporating a second TIB downstream of each qubit/cavity that switches in tandem with its upstream partner (see Fig.~1(b)). In a compact layout (connections much less than a wavelength), this would again make the other $N - 1$ channels appear as open circuits and direct power out the microwave receiver.

\section{Expected Infidelity with Binary Signaling}

The solid lines plotted in Fig.~4(b) in the main text are 
predictions of a Gaussian noise model with no free parameters, a standard method for calculating error probabilities with binary signaling~\cite{Smadhow:2014}.  
We reconstruct the $k^\textrm{th}$ digital bit of the $q^\textrm{th}$ channel from the sign (polarity) of the reconstructed analog bit $A^r_{q,k}$.  The efficiency of this process is determined by the distribution $f_q$ from which $A^r_{q,k}$ is drawn.  The model assumes that the analog voltage $A^r_{q,k}$ (calculated from Eq.~5 in the main text) has a Gaussian distribution over many repeated transmission events:
\begin{eqnarray}
f_q = \frac{1}{\sigma_q \sqrt{2 \pi}}\textrm{exp}\left(-\frac{(x-\mu_q)^2}{2 \sigma_q^2}\right).
\end{eqnarray}
Consequently, the predicted infidelity $I_q$ depends only on the mean $\mu_q$ and variance $\sigma_q^2$ of $A^r_{q,k}$:
\begin{eqnarray}
I_q &=& \int_{-\infty}^0 {f_q(x) dx}, \nonumber \\
&=& \frac{1}{2} \textrm{erfc}\left(\frac{\mu_q}{2\sigma_q}\right), 
\label{InGen}
\end{eqnarray}
with erfc the complementary error function~\cite{abramowitz:1964}.

We assume that the noise $v(t)$ in our measurement channel is white, and satisfies
\begin{eqnarray}
\label{noiseassumptions}
\langle v(t) \rangle_t &=& 0, \\ \nonumber
\langle v(t) v(t-t') \rangle_t &=& Z_0 S \delta(t').
\end{eqnarray}
Here the subscripts indicate a time average, and $S$ is the spectral density of the noise measured with a spectrum analyzer with no signal transmitted in any channel.  We may then account for this noise by addding $v(t)$ to the time trace acquired during the transmission of the $k^\textrm{th}$ bit (Eq.~4 in the main text):

\begin{eqnarray}
V_k(t) \to A_{1,k} W_n(t) + A_{2,k} W_m(t) + v(t).
\end{eqnarray}

The desired moments of $f_q$ can then be calculated in the standard way.  For brevity, we now set $q=k=1$ and drop the bit subscript:
\begin{eqnarray}
\label{un}
\mu_1 &=& \langle A^r_{1} \rangle \nonumber \\
&=& \frac{1}{t_w} \biggr\langle \int_0^{t_w} \Big[|T_{\textrm{ch}1}| W_n(t) A_1 + |T_{\textrm{ch}2}| W_m(t) A_2 + v(t)\Big] W_n(t) dt \biggr\rangle \nonumber \\
&=& |T_{\textrm{ch}1}| A_1 U.
\end{eqnarray}
Here $\langle \cdot \rangle$ indicates an ensemble average, 
\begin{eqnarray}
U \equiv \frac{1}{t_w} \int_0^{t_w} W_n(t')^2dt',
\end{eqnarray}
and $T_{\textrm{ch}1}$ ($T_{\textrm{ch}2}$) is the transmission coefficient of the TIB in channel 1 (2).  We intentionally separate this factor from our finite-bandwidth realizations of the Walsh codes so that their maximal and minimal values are 1 and -1, just like the elements of $\{w_n\}$. In the high-bandwidth limit, $\{W_n\}$ approaches $\{w_n\}$ and the Walsh code normalization condition
\begin{eqnarray}
\frac{1}{t_w} \int_0^{t_w} w_n(t')^2dt' = 1,
\end{eqnarray}
makes $U = 1$.  

To measure $|T_{\textrm{ch}1}| A_1$ directly, we transmit a stream of $1$'s through the first channel, (no signal is transmitted through the other channel), and measure the spectral density of the transmission peak $S_1$, which provides
\begin{eqnarray}
\label{T1}
|T_{\textrm{ch}1}| A_1 = \sqrt{Z_0 S_1 B_w},
\end{eqnarray}
with the resolution bandwidth of the spectrum analyzer $B_w$.

To measure the second moment, we evaluate
\begin{eqnarray}
\langle \left(A^r_{1}\right)^2\rangle &=&  \biggr\langle \frac{1}{t_w^2} \int_0^{t_w} \int_0^{t_w} \Big[ v(t') + |T_{\textrm{ch}1}| W_n(t') A_1 + |T_{\textrm{ch}2}| W_m(t') A_2 \Big] W_n(t') \nonumber \\
&& \times \Big[ v(t'') + |T_{\textrm{ch}1}| W_n(t'') A_1 + |T_{\textrm{ch}2}| W_m(t'') A_2 \Big] W_n(t'')dt'dt''\biggr\rangle \nonumber \\
&=&  \frac{S Z_0 U}{t_w}  +  |T_{\textrm{ch}1}|^2 A_1^2  U^2.
\end{eqnarray}
Here in the last line we have used both assumptions in Eq.~\ref{noiseassumptions}, the fact that $v(t)$ and $W_n(t)$ are uncorrelated, and the orthogonality of the $\{W_n\}$.

The variance may then be computed as
\begin{eqnarray}
\label{sigma}
\sigma_1^2 &=& \langle \left(A^r_{1}\right)^2\rangle - \langle A^r_{1} \rangle^2 \nonumber \\
&=& \frac{S Z_0 U}{t_w}.
\end{eqnarray}

Substituting the mean $\mu_1$ from Eqs.~\ref{un} and~\ref{T1} and the standard deviation $\sigma_1$ from Eq.~\ref{sigma} into Eq.~\ref{InGen}, we obtain the desired prediction for the infidelity of the first channel:

\begin{eqnarray}
\label{InfidelityPred}
I_1 &=& \frac{1}{2} \textrm{erfc}\left(\sqrt{\frac{S_1 B_w t_w U}{2S}}\right).
\end{eqnarray}

The effect of an attenuated signal is then easily accounted for by scaling the signal power $S_1 B_w$ in Eq.~\ref{InfidelityPred} by the attenuation constant.  Note that $U$'s deviation from 1 is approximated by three times the ratio of the switching time $\tau$ to the Walsh period $t_w$, as the signal is not fully transmitted while the TIBs switch between the transmit and invert operation modes~\footnote{The factor of 3 appears because the Walsh codes used in our demonstration switch on average 3 times per Walsh period.}.  This reduces signal power by 11\% when $t_w = 400$ ns.  The effect is a factor of 5 (25) smaller when $t_w$ = 2 (10) $\mu$s.

The predictions plotted in Fig.~4(b) of the main text are obtained with Eq.~\ref{InfidelityPred}, using the measured input parameters shown in Tab.~\ref{tab:thetab}.

\begin{table}[hbt]
\centering
\caption{Input parameters for the Gaussian noise model of readout infidelity, plotted in Fig.~4(b) of the main text.}
\label{tab:thetab}
\begin{tabular}{|l|l|l|l|l|l|}
\hline
$S$       & $S_1$    & $S_2$    & $B_w$   & $t_w$ & $\tau$                            \\ \hline
-90.4 dBm & -50.1 dBm & -50.5 dBm & 50 KHz & 10 $\mu$s, 2 $\mu$s, or 400 ns & 15 ns\\ \hline
\end{tabular}
\end{table}

\end{document}